\documentclass[runningheads]{llncs}
\usepackage{graphicx}
\usepackage{color}
\usepackage{amsmath, amssymb}
\usepackage[caption=false,font=footnotesize]{subfig}
\begin{document}

\title{Network Classification in Temporal Networks Using Motifs\thanks{This work was supported in part by  U.S. Army Research Laboratory and the U.K. Ministry of Defence under
Agreement Number W911NF-16-3-0001, ARL Cooperative Agreement W911NF-09-2-0053. The views and conclusions contained in this document are those of the authors and should not be interpreted as representing the official policies, either expressed or implied, of the U.S. Army Research Laboratory, the U.S. Government, the UK Ministry of Defence or the UK Government. The U.S. and UK Governments are authorized to reproduce and distribute reprints for Government purposes notwithstanding any copyright notation hereon.} }

%
%
\author{Kun Tu\inst{1}\and Jian Li\inst{1}\and Don Towsley\inst{1}\and Dave Braines\inst{2}\and Liam D. Turner\inst{3}}
\authorrunning{K. Tu  et al.}
%
\institute{$^1$University of Massachusetts Amherst, $^2$IBM UK, $^3$Cardiff University}

\maketitle              
\vspace{-0.2in}
\begin{abstract}

Network classification has a variety of applications, such as detecting communities within networks and finding similarities between those representing different aspects of the real world. However, most existing work in this area focus on examining static undirected networks without considering directed edges or temporality. In this paper, we propose a new methodology that utilizes feature representation for network classification based on the temporal motif distribution of the network and a null model for comparing against random graphs. Experimental results show that our method improves accuracy by up $10\%$ compared to the state-of-the-art embedding method in network classification, for tasks such as classifying network type, identifying communities in email exchange network, and identifying users given their app-switching behaviors.

\end{abstract}

\vspace{-0.4in}
\section{Introduction}\label{sec:intro}

Networks, where interacting elements are denoted as nodes and interactions are denoted as edges, are a fundamental tool to study complex systems \cite{albert02,newman10}, including social, communication, biology and economics networks.  Typical analysis usually models these systems as static graphs that describe the relations between nodes.  However, in many realistic applications, the relations are often not fixed but instead evolving over time \cite{holme12,kovanen11,paranjape17}.  Modeling these \emph{temporal} properties is of additional interest as it provides a richer characterization of relations between nodes in the network.  

When considering dynamic changes to a network, most studies aggregate temporal information into a sequence of snapshots, with each representing the graph over some time period \cite{araujo14,dunlavy11,tantipathananandh07}. However this limits the ability to capture changes at a finer granularity, and hence loses the richness of the temporal information contained in the data \cite{paranjape17}. In this paper, we incorporate temporality at the edge level, by examining \emph{temporal networks}\cite{paranjape17}, which can be defined as a set of nodes and a collection of directed temporal edges with attributed timestamps. For example, an email network can be represented as a sequence of timestamped directed edges, where each represents an email sent from one person to another. 

Examining temporal patterns provides a useful basis to explore insights into how the nodes in a network interact, as this incorporates both network topology and dynamics. For example, in complex network classification problems, the underlying network topology for the community has an impact on the development of the community since the interactions between nodes can be treated as backbones of the community.  Thus, we can identify a community by comparing and distinguishing its connected modes. However, most of the current studies in this area focus on examining a single complex network (e.g. \cite{albert02}), with little direct attention on examining how temporal patterns can be used to identify and classify networks, or provide a basis for examining similarity and clustering between networks.

In this paper, we propose a new network embedding (feature representation) methodology for network classification in temporal networks based on a network's \emph{temporal network motif distribution}. Network motifs capture the local structure of a network through summarising whether small patterns in the network occur significantly more or less frequently than random graphs (using a null model). In temporal networks, \emph{temporal network motifs} are defined as induced subgraphs on sequences of temporal edges \cite{paranjape17}.  In particular, we consider two sub-problems (i) how temporal network motifs can be used to classify the \emph{network type} (defined as the context of the relation or interaction between nodes, such as social ties in social networks); and (ii) identification of a particular network from its temporal topological structure within a period.   


We first propose a generic framework to construct vectors for \emph{feature representations} of temporal directed graphs from their topological structure using temporal motif distribution and null models. We argue that this fixed length feature representation can be used to classify and compare networks of varying sizes and period with high accuracy. We apply various well-known machine learning models along with our graph feature representation for these two network classifications, and make a comparison with the state-of-the-art method,  \emph{struc2vec} \cite{ribeiro17}. Our results show that the motif-based feature representation models can significantly outperform struc2vec.  Furthermore, we observe that temporal information improves the accuracy of network identification in comparison to considering the network as a static graph, with a combination of static and temporal features bringing further improvement; providing a basis to reflect on the benefits and drawables of considering a network as a static or temporal graph.

\section{Related Work}\label{sec:related}
The primary focus of related works in classifying networks involves examining the topological structure of the graph. For example, kernel methods have been used to calculate similarities between static undirected graphs \cite{gauzere2015treelet}, \cite{yanardag15}.  
 However, the corresponding computational complexity grows significantly with the increase of network size.  Different node embedding techniques have been proposed in the past years, such as node2Vec \cite{grover16}, DeepWalk \cite{perozzi14}, Line \cite{tang15} and Local Linear Embedding \cite{roweis00} that use feature vectors to embed nodes into high-dimensional space and empirically perform well.  However, these methods can only be applied to tasks, e.g., classification of nodes but not on the whole networks.   

Additionally, several approaches have been proposed to aggregate node feature vectors to a feature vector for networks.  For example,  graph-coarsening approach \cite{defferrard2016convolutional} computes a hierarchical structure containing multiple layers, nodes in lower layers are clustered and combined as node in upper layers using element-wise max-pooling.   However, this has high computational complexity.  A recent paper \cite{mellor18} applied motif-based feature representation for clustering network types, however, it is not clear how motifs can be used to identify communities within the same network.  Some approaches \cite{niepert2016learning} 
 define an order of nodes and concatenate their feature vectors for a convolutional neural network for classification, however, this can only be applied to undirected static networks.  Most recently,  \emph{struc2vec} \cite{ribeiro17} was proposed with {meanfield} and loopy belief propagation \cite{murphy1999loopy} to aggregate node embedding to graph representation and empirically shown to outperform previous approaches. Given this, we use struc2vec as a baseline for comparison in this paper.

\section{Problem Formulation}
\begin{figure}
\vspace{-0.1in}
	\centering
	\subfloat[Static Motifs]{
		\includegraphics[width=0.43\linewidth]{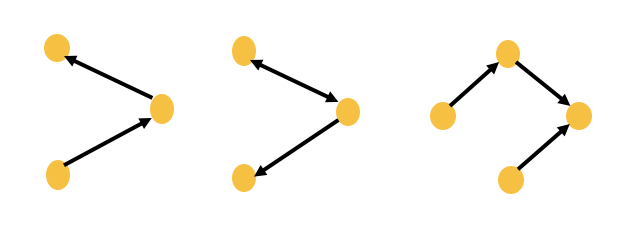}
	}
	\subfloat[Temporal Motifs]{
			\includegraphics[width=0.43\linewidth]{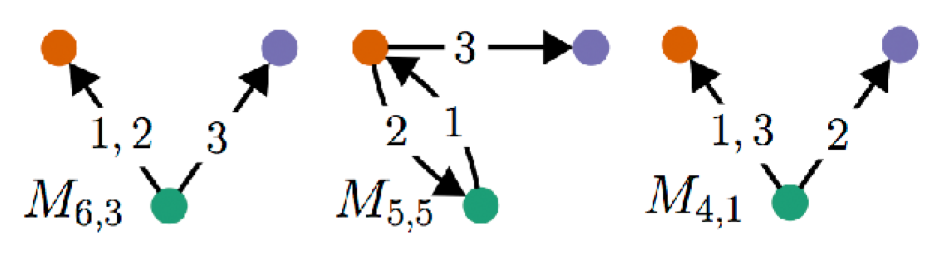}
			}
	\caption{Network Motifs. (a) example of triad (three-node motif) and tetrad (four-node motif); (b) temporal motifs whose edges appear in a specific order (refer to \cite{paranjape17} for more details.}
	\label{fig:motifs}
	\vspace{-0.2in}
\end{figure}

We use definitions of temporal networks and temporal network motifs as defined in \cite{paranjape17}.  Due to space constraints, we give some examples on static and temporal motifs in Figure~\ref{fig:motifs}, and refer interesting readers to \cite{paranjape17} for further detail. 

%
%

%
%

\textbf{Problem formulation} The task of network type classification or network identification can be formalized as follow:

Denote $\{G_i(V_i, E_i, L_i)\}_{i=1}^N$ as a set of $N$ (sub)graphs, where $V_i$ is a set of nodes and $E_i$ is a set of timestamped edges in $G_i$. Suppose that graphs can be categorized into $D$ classes, where $D<N$.  We associate each graph $G_i$ with a label $L_i\in\{1,\cdots, D\}$.

Let $f: \{G_i\}\rightarrow \mathbb{R}^m$ be a \emph{mapping function} (also called graph embedding function) from $G_i$ to a $1\times m$ \emph{feature representation vector} that is defined using subgraph ratio profiles (SRP) of temporal or static motifs.  (We will formally define SRP in Section~\ref{sec:embedding}).

Let $g: \mathbb{R}^m \rightarrow P\in \mathbb{R}^D$ be \emph{classifiers} that map a feature representation to a categorical distribution $P$ for $D$ labels.  Thus the probability distribution of $G_i$'s label can be represented as $P_i=[p_{i,1}, \dots, p_{i,D}] = g(f(G_i))$. 
 
 Our goal is to solve this classification problem through designing a embedding function  $f$ and selecting a machine learning model $g$ to minimize \emph{the sum of cross entropy \cite{de05} for all graphs}
\begin{equation}
\arg\min_{g,f} [-\sum_{i}\sum_{j=1}^D\mathbf{1}_{L_i=j} \log(p_{i,j})]
= \arg\min_{g,f} [-\sum_{i}\log(p_{i,L_i})].
\end{equation}
We obtain $g$ by training machine learning models.  In next section, we discuss how to design an embedding function $f$ for temporal networks using motifs.

\section{Network Embedding Using Motifs}\label{sec:embedding}

As discussed in Section \ref{sec:related}, network embedding has received a lot of attentions due to its effect on the performance of network classification.  However, previous works have primarily focused on examining this with static networks. Applying these techniques directly to temporal networks loses temporal information and may result in poor accuracy. Therefore, we introduce a new temporal network embedding technique that uses temporal network motifs. 

A temporal embedding needs to be independent of network size and the time period the network covers. While previous works have shown that the counting and probability distribution of motifs are strongly related to network types \cite{paranjape17}, motif counts may be different across networks. 
Therefore, we use subgraph ratio profiles (SRP) for temporal network embedding, which is computed using motif counts from both the network in question and random graphs produced using null models.
\begin{definition}
A null model is a generative model that generates random graphs that matches an specific graph in some of its structural features such as the degrees of nodes or number of nodes and edges \cite{newman04}.
\end{definition}
In statistics, random graphs in a null model are used for comparison to verify if the graph in question displays a certain feature. In our study, we use a null model to compare the counts of motifs in a network against random graphs. We consider ensembles of time-shuffled data for null model in temporal networks.  The difference between these is then used to construct an SRP as a feature representation of the network.

\begin{definition}
Subgraph ratio profile (SRP) \cite{milo04} for a motif $i$ is defined  as 
\begin{equation}
SRP_i=\frac{\Delta_i}{\sqrt{\sum \Delta_i^2}},
\end{equation}
where $\Delta_i$ is a normalized term that measure the difference between the count of motif $i$ in an empirical network (denoted as $N_{observed_i}$) and the average count in random networks in a null model (denoted as $<N_{random_i}>$):
\begin{equation}
\Delta_i=\frac{N_{observed_i}-<N_{random_i}>}{N_{observed_i}+<N_{random_i}>+\epsilon},
\end{equation}
where 
$\epsilon$ (usually set to $4$) is an error term to make sure that $\Delta_i$ is not too large when the motif rarely appears in both the empirical and random graphs.
\end{definition}
Since an SRP for a motif is a normalized term, it can compare networks of different sizes.  For static directed networks, we consider the null model for random graphs with the same number of nodes and edges. The network embedding is a vector containing $16$ SRPs for static triads. For null models of temporal directed networks, we further randomly the order of temporal edges. The embedding contains the SRPs for $36$ temporal motifs (Figure $3$ in \cite{paranjape17}).



\section{Experiments} 
We use two types of real-world temporal networks as the basis for this study: email exchanges within different departments in an European institution and the app switching behavior of 53 smartphone users over 42 days. To evaluate the effectiveness in our temporal network embedding in network classification, we apply it to several machine learning models on network classification based on topological network structure and compare its performance to embeddings using static motifs and other techniques.  More specifically, we are interested in classifying the network type and identifying the department given the topological structure of an email network.  We also attempt to identify a person given their mobile app-switching behavior represented as a temporal network.

\vspace{-0.1in}
\subsection{Datasets}
{\textbf{EmailEU} \cite{Yin17} is a directed temporal network constructed from email exchanges in a large European research institution for a 803-day period.}  It contains $986$ email addresses as nodes and $332,334$ emails as edges with timestamps.  There are $42$ ground truth departments in the dataset and we constructed subgraphs from these departments with size larger than $10$.

{\textbf{SwitchApp} (from the tymer project \cite{noe2017timing}) contains application switching data for $53$ Android users for a $42$-day period.}  We construct a directed temporal network for each user, where a directed edge (denoted as $e_{uv}$) with an integer {timestamp} $t$ represents a user switching from an app $u$ to another $v$ at time $t$. 

\vspace{-0.1in}
\subsection{Experiment Setup} 
We compute SRPs for both temporal and static motifs for the EmailEU and SwitchApp temporal networks.  {Networks in EmailEU and SwitchApp are labeled with $0$ and $1$, respectively.}  We consider four widely used machine learning models that have good performance with small amount of training data:   {XGBoosting \cite{chen16}, SVM \cite{cortes1995support}, random forest \cite{svetnik03} and AdaBoost \cite{Ratsch01}.}  We use {grid search method} to search the best hyper-parameters for these models. {$10$-fold cross-validation} is adopted to split the data for selected models with the best parameter:  For XGBoosting algorithm, the learning rate is set to $0.1$, maximal tree depth is set to $8$, minimal child weight is $1$ and the subsample ratio of train instances is set to $0.8$. The regularization weight in SVM is set to $2$. In random forest, the number of tree is set to $400$ and the minimal samples required to split a tree node is $2$. 

We also make a comparison of our temporal embedding with the state-of-the-art method, {\emph{struc2vec} \cite{ribeiro17}}, in network classification. Since struc2vec requires node attributes for network embeddings, we compute the in/out degree, betweenness, closeness and in/out degree centrality for each node. The length of network embedding is decided using grid search and 10-fold cross-validation.

\vspace{-0.1in}
\subsection{Network Types Classification}
We first testify if temporal motifs provide more information than static motifs for network type classification.  

\begin{figure}[h]
\vspace{-0.1in}
	\centering
	\includegraphics[width=0.5\linewidth]{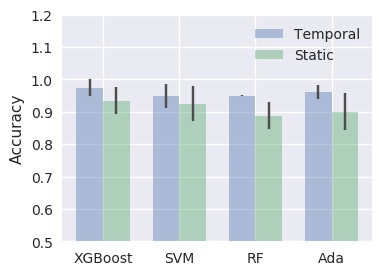}
	\caption{Classifying EmailEU and SwitchApp Temporal Networks.}
	\label{fig:networkType}
	\vspace{-0.2in}
\end{figure}

{From Figure~\ref{fig:networkType}, we observe that temporal information improve the network type classification in all models considered here. }

\vspace{-0.1in}
\subsection{Temporal Network Identification}
{Additionally, we examine if individual networks can be identified from their structure. In the EmailEU dataset, we attempt to identify which department the emails belong to. For the SwitchApp dataset, we attempt to identify a particular user given his daily app-switching behaviors represented as temporal networks.}

{For the EmailEU dataset, multiple temporal and static networks are constructed for each departments from email exchanges.  For the SwitchApp dataset, $42$ temporal and static networks are generated for each person from his app switching behaviors every day.  XGBoosting, SVM, random forest and AdaBoosting are implemented using five different network feature representations: subgraph ratio profile (SRP) with temporal (``Temporal'') and with static (``Static") motifs, concatenated SRPs with both temporal and static motif (``Temp+Static") and struc2vec representation (``S2V"). 

\begin{figure*}[ht]
\centering
\begin{minipage}{.5\textwidth}
\centering
\includegraphics[width=1\linewidth]{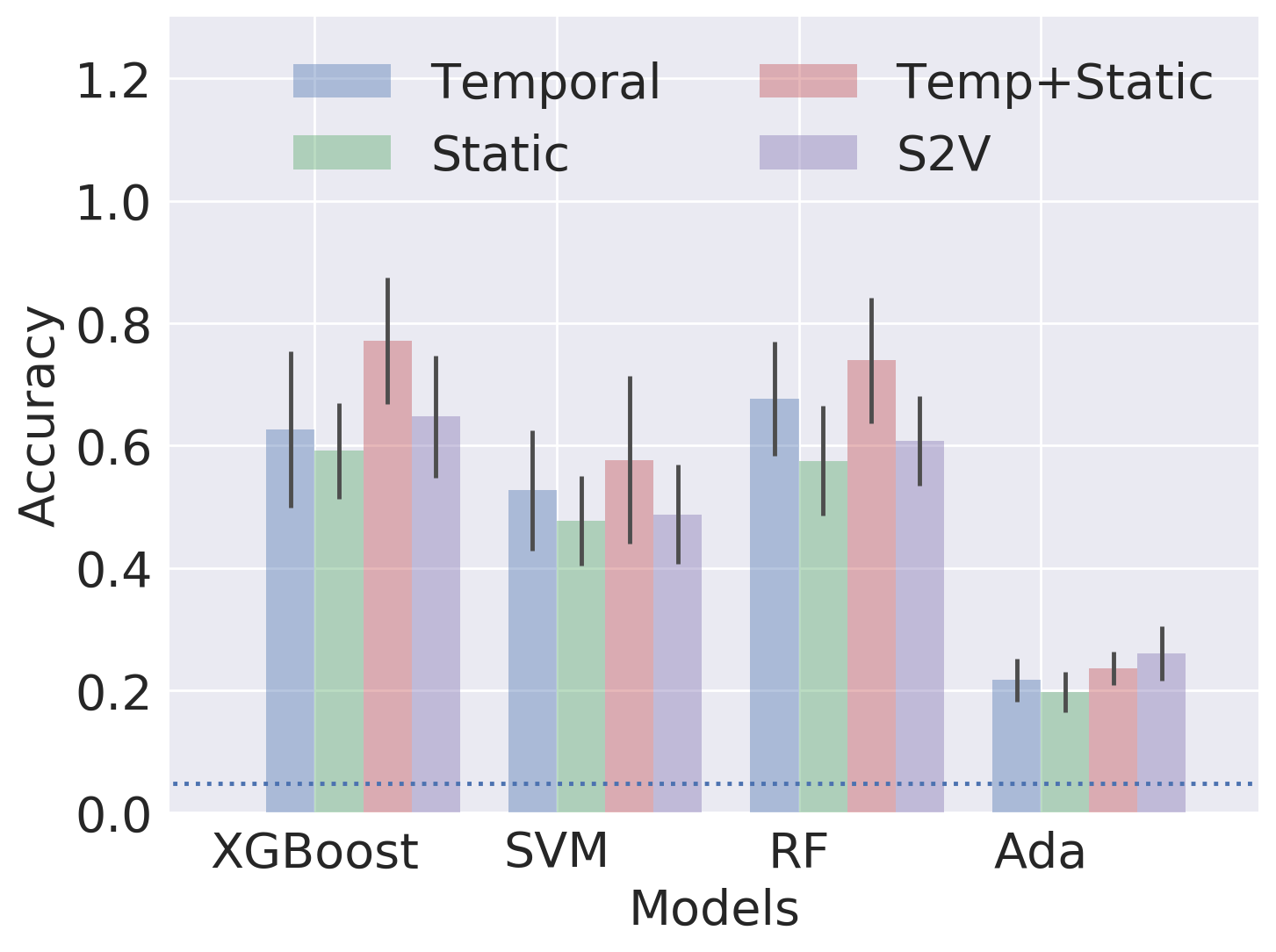}
\end{minipage}\hfill
\begin{minipage}{.5\textwidth}
\centering
\includegraphics[width=1\linewidth]{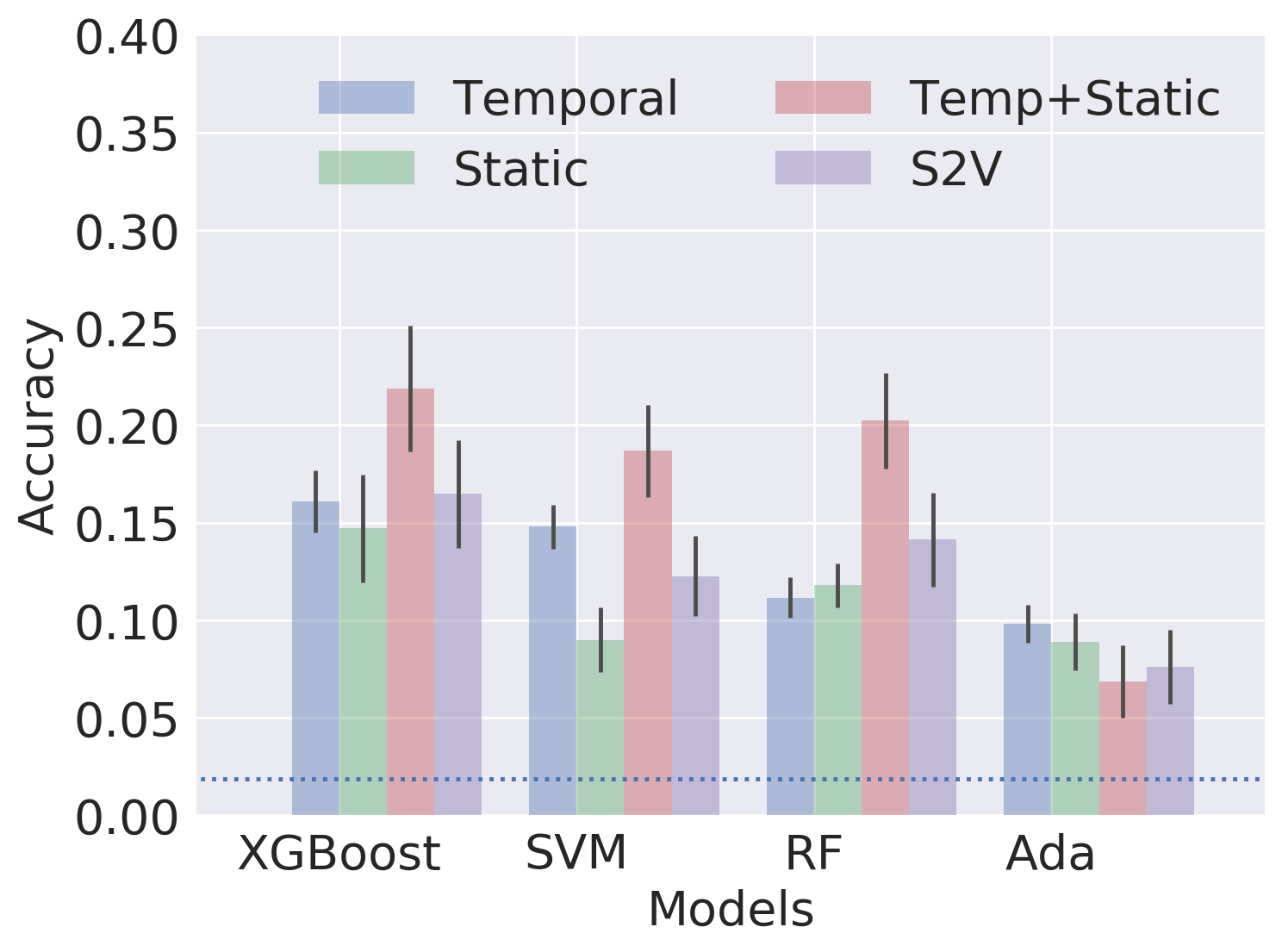}
\end{minipage}
\caption{{\bf \emph{(Left):}} Department Identification in EmailEU-Core dataset. 
{\bf \emph{(Right):}} User Identification in SwitchApp Temporal Networks.  }
\label{fig:departmentIdentification}
\vspace{-0.2in}
\end{figure*}

{The results for EmailEU and SwitchApp are shown in Figure~\ref{fig:departmentIdentification}. The dash line is the accuracy of a random selection model. The accuracy achieved by temporal motifs embedding is slightly better than that of static motifs embedding. However, there exists a significant improvement with concatenated motif features (``Temp+Static"), which suggests that both temporal and static motifs are useful for network identification (of departments or personal app switching behavior).   Furthermore, at least one of our motif-based network embeddings can outperform the state-of-the-art algorithm, \emph{struc2vec}.} The best machine learning model is XGBoost with both temporal and static motif embedding. 

\section{Conclusion}
We propose a network embedding using motifs and null models to classify networks based on their topological structure. Experiments with real-world datasets show that both temporal and static motifs are important to network type classification and network identification. Therefore, concatenating these two embedding yields better accuracy and outperform the state-of-the-art method.

\bibliographystyle{splncs04}
\bibliography{refs}
%
%
%
%
%
%
%
%
\end{document}